\documentclass[%
 reprint,
superscriptaddress,
showpacs,
nobibnotes,
 aps,
 prl
]{revtex4-1}

\usepackage{graphicx}% Include figure files
\usepackage{dcolumn}% Align table columns on decimal point
\usepackage{bm}% bold math
\usepackage{amsfonts}
\usepackage{amsmath}
\begin{document}

%\preprint{APS/123-QED}

\title{Efficiently controllable graphs}% Force line breaks with \\
%\thanks{A footnote to the article title}% %
\author{Can Gokler}
%\email{gokler@fas.harvard.edu} 
\affiliation{ School of Engineering and Applied Sciences, Harvard University, Cambridge MA 02138, USA}
%\altaffiliation{ Research Laboratory for Electronics, Massachusetts Institute of Technology, Cambridge MA 02139, USA }
 %\altaffiliation[Also at ]{...}%Lines break automatically or can be forced with \\

%\affiliation{%

%\collaboration{...}%\noaffiliation%

 %\homepage{....}
%\affiliation{
%\affiliation{

\author{Seth Lloyd}%
\affiliation{Department of Mechanical Engineering, Massachusetts Institute of Technology, Cambridge MA 02139, USA 
}

\author{Peter Shor}
\affiliation{Department of Mathematics, Massachusetts Institute of Technology, Cambridge MA 02139, USA}
%\affiliation{%


\date{\today}% It is always \today, today,

\author{Kevin Thompson}
\affiliation{ School of Engineering and Applied Sciences, Harvard University, Cambridge MA 02138, USA}

%\altaffiliation{ Research Laboratory for Electronics, Massachusetts Institute of Technology, Cambridge MA 02139, USA}

\begin{abstract}
We investigate graphs that can be disconnected into small
components by removing a vanishingly small fraction of their vertices.
We show that when a quantum network is described by such a graph,
the network is efficiently controllable, 
in the sense that universal quantum computation can be performed 
using a control sequence polynomial in the size of the network 
while controlling a vanishingly small fraction of subsystems. 
We show that networks corresponding to finite-dimensional lattices
are efficently controllable, and explore generalizations to percolation 
clusters and random graphs. We show that the classical computational complexity of estimating the ground state of Hamiltonians described by controllable graphs
is polynomial in the number of subsystems/qubits.

%\begin{description}
%\end{description}
\end{abstract}


                             
%02.30.Yy	Control theory  
%03.67.-a	Quantum information   
%03.65.Aa	Quantum systems with finite Hilbert space
%03.65.-w	Quantum mechanics 
%03.67.Lx	Quantum computation architectures and implementations
%07.05.Dz	Control systems             
    
\maketitle

%\tableofcontents

%\section{First section}

Controlling large quantum networks and performing universal quantum 
computation are two important and related problems in quantum information
processing.  A common goal is to perform control and computation
efficiently, by accessing a minimum number of directly controlled parts. 
Quantum networks were introduced in \cite{Deutsch}.  In \cite{Lloyd2004}
it was shown that almost any quantum network with single probe is 
controllable.  If controls can be applied to quantum degrees of freedom
in a pairwise fashion, then the control is
computationally universal \cite{Lloyd1995}.  The connectivity of 
the graph of interactions plays an important role in controllability
and computation \cite{Altafini}. 
Under mild assumptions about network topology and the algebra of controls,
\cite{Burgarth2009} gave sufficient conditions for a network to be
controlled using a small number of control qubits, without regard
for the efficiency of the control sequence.
See \cite{Barabasi2011} for a similar study of classical linear systems. 
In suitable systems, quantum computation is possible with only a few 
control qubits \cite{Burgarth, Kay}. As suggested in these papers, 
spin chains with specific Hamiltonians can give controllability as well as 
the ability to enact efficient universal quantum computation on the chain. These results raise the question of when it is possible to perform
universal quantum control and quantum computation efficiently on a
general quantum network. This paper shows in a general setting
that it is possible to perform universal quantum control
and computation in time polynomial in network size on a wide variety of quantum networks while acting on only a vanishingly small fraction of their nodes. This naturally leads one to define an interesting class of graphs, which we call {\it{efficiently controllable graphs}}, which admit efficient control by acting on a vanishingly small fraction of controlled nodes. Existence, construction and analysis of this new class of graphs pose an intriguing problem in graph theory. In this work, we construct several examples of such families of graphs and show that the ground states of Hamiltonians of systems whose interactions are determined by such graphs can be approximated efficiently.

Consider a quantum system consisting of subsystems interacting via
local Hamiltonians. 
The subsystems can be represented as vertices of an interaction (hyper)graph 
where (hyper)edges exist only between vertices corresponding to coupled
subsystems.  For simplicity of exposition, we will restrict our attention
to pairwise Hamiltonians and interaction graphs.   However, all our results
apply to general local Hamiltonians and interaction hypergraphs.
Without loss of generality, here the 
quantum systems are restricted to networks of qubits.

Take a single spin which is promoted to be controlled and observed 
on its own, a universal quantum interface \cite{Lloyd2004}. The ability 
to act by any unitary operation and a single measurement translates into 
universal quantum computation on that qubit. Together with the 
interaction of the qubit with the rest of the graph, one may perform 
universal quantum control on the whole system. With a single interface, 
the number of quantum operations required to approximate any unitary 
operation to a fixed accuracy grows exponentially with the number of 
spins in the system. As will be shown, by promoting more qubits to be 
interfaces, controlled spins, a polynomial growth can be reached as 
required for a scalable quantum computer architecture.  

Since implementation becomes more complex as the number of controlled spins grows, 
a scalable implementation needs to choose the smallest
number of interfaces possible while still preserving polynomial 
efficiency of quantum computation. The primary purpose of this article 
is to show that there exist families of interaction graphs 
such that the quantum computational efficiency scales polynomially 
with the number of vertices and the fraction of controlled qubits 
approaches zero as the number of nodes in the graph goes to infinity. That is,
there are scalable and efficient quantum computer architecture schemes 
that make use of vanishing fraction of controlled qubits. Lattices and 
uniform tilings are examples of such families. Expanders and complete graphs 
are not likely to be such graphs. Not every family of graphs admits such 
schemes, therefore we define a new family, which we call 
efficently controllable
graphs.  An efficiently controllable graph is a graph
that can be divided into components of size $\text{poly}(\log(n))$
by removing a vanishingly small fraction of vertices where $n$ is the number of vertices.  Assuming controllability conditions, we prove
that on a quantum network described by such a graph one can perform
universal quantum computation efficiently 
by controlling a vanishingly small fraction of vertices in
the limit that the size of the graph goes to infinity 

%Consider a connected network of $n$ spins, $c$ of which are controls. 
%Call the subsystem formed by uncontrolled qubits $S$ and the subsystem of
%controlled qubits $C$. Consider a time independent coupling Hamiltonian 
%$H=H_S+ \sum_{j \in C} A_x^j 
%\otimes \sigma_x^j + A_y^j \otimes \sigma_y^j + A_z^j \otimes \sigma_z^j + 
%H_C$ where $H_S$ acts on S, $H_C$ acts on C and $A_{\alpha}^j$ are 
%Hermitian operators acting on $S$. Assume that, on $C$, one can
%perform arbitrary single qubit Hamiltonians $h^j$ and can
%measure each qubit with respect to one basis. The Lie algebra spanned by 
%$ \{ H, h^j \}$ is the algebra of the whole network for almost 
%any Hamiltonian H. The system is then coherently controllable and 
%any unitary operation can be applied to it. Here, we assume that by controlling a 
%small number of additional qubits, full controllability can be achieved. 
%See for example \cite{Burgarth2009, Burgarth13, Godsil10} for a sufficient condition for 
%local controllability.

%The ability to perform arbitrary single qubit measurements on $C$ translates into the 
%ability to effect arbitrary generalized open-system transformations on 
%the network; in 
%addition, if $S$ decomposes into non-interacting parts, $S_1$ 
%and $S_2,$ each interacting with $C$, then $C$ can be also used 
%as a quantum communication channel between the parts; finally,
%if the control Hamiltonian can
%be varied arbitrarily rapidly, then the effective Hamiltonian
%of $C$ can be set to zero, decoupling $S_1$ and $S_2$
%\cite{Lloyd2004}. 

In a connected network architecture of $n$ spins satisfying certain assumptions on the drift and control Hamiltonians, enacting 
an arbitrary unitary operation within a constant error $\epsilon$ 
requires $\text{s}(n)=O(2^{nx} \text{poly}( 1 / \epsilon))$ elementary operations for some $x$ as we show now. Consider a connected network of $n$ spins. To make the argument simple, restrict to the Hamiltonian with a single control term $H(t)=H_0 + H_c \gamma(t)$ where $-iH_0, -iH_c \in su(d=2^n)$ are bounded. We assume that the pair $(H_0, H_c)$ is controllable. See for example \cite{Altafini, Burgarth2009, Burgarth13, Godsil10} for sufficient conditions for 
local controllability. The control problem is to find $\gamma(t)$ to drive an initial unitary, which is the identity, to the $\epsilon$ neighborhood of a final unitary. The Hamiltonian defines a flow in the set of unitaries as $\dot{U}(t)=-i(H_0 + H_c \gamma(t))U(t)$ where $U(t) \in SU(d)$. Assume that $H_0=\sum_k E_k P_k$ is non-degenerate (if $H_0$ is degenerate, one can break the degeneracy by applying a constant control $\lambda H_c$\cite{note3}) to accuracy $1 / d^r $ where $P_k=|k\rangle \langle k|$ is the projector onto the eigenvector $|k\rangle$ of $H_0$ with eigenvalue $E_k$. Assume also that $|\Delta_{jk}|=|E_j-E_k| > 1/d^r$ are distinct and $P_kH_cP_m\neq 0$ for all $k$, $m$ (this condition can be relaxed: see \cite{note1}), and $||H_c||=O(1)$ where $||U||=\sup_{y \in \mathbb{C}^d, y^{\dagger}y=1}||Uy||$ is the operator norm. Drive the system with control with amplitude $A$, resonant frequency $\Delta_{jm}$ and phase $\phi$, so that the Hamiltonian takes the form $H(t)=H_0+A\cos(\Delta_{jm}t+\phi)H_c$. Now go the interaction picture by defining $U_i(t)$ via $U(t)=e^{-i t H_0}U_i(t)$. Then 

\noindent
\begin{equation*}
\dot{U_i}(t)=-ie^{i t H_0}A\cos(\Delta_{jm}t+\phi)H_c e^{-i t H_0}U_i(t).
\end{equation*}

The approximate solution of this equation is given by the Magnus expansion \cite{Blanes} as

\noindent
\begin{equation*}
\begin{split}
 U_0(T) & = \exp(-i \Omega(T)) \\ 
& =e^{-i \int_0^Tdt\exp(i t H_0)A\cos(\Delta_{jm}t+\phi)H_c\exp(-i t H_0)}
\end{split}
\end{equation*}

 \noindent
 with error $|| U_0(T) - U_i(T)|| = O(|A|^2T^2||H_c||^2 )$ where $|A| T ||H_c|| < \pi$ for the convergence of the series. We write $\Omega(T)=\Omega_1(T)+\Omega_2(T)$ as the sum of the resonant term and the off-resonant term. The resonant term is given by
 
 \noindent
\begin{equation*}
\Omega_1(T)=\frac{A}{2}T(e^{-i \phi}P_j H_c P_m + e^{i\phi} P_m H_c P_j)
\end{equation*}

\noindent
and $||\Omega_2(T)|| = O(|A| d^{2+r} ||H_c|| ) $. The error in neglecting $\Omega_2(T)$ is given by \cite{KitaevVyalyi}

\noindent
\begin{equation*}
\begin{split}
 & || U_0(T) - e^{-i\Omega_1(T)}|| \\
 & \leq || e^{-i(\Omega_1(T)+\Omega_2(T))} - e^{-i\Omega_1(T)}e^{-i\Omega_2(T)}|| \\ 
 & \indent +||e^{-i\Omega_1(T)}e^{-i\Omega_2(T)}- e^{-i\Omega_1(T)}|| \\
 & = O(||\Omega_1(T)|| ||\Omega_2(T) || +||\Omega_2(T) ||) \\ 
& = O(|A|^2Td^{2+r} ||H_c||^2+|A|d^{2+r}||H_c||).
 \end{split}
\end{equation*}

Now bound the error between $U_i(T)$ and $e^{-i\Omega_1(T)}$ using the triangle inequality as

\noindent
\begin{equation*}
\begin{split}
&||U_i(T)-e^{-i\Omega_1(T)}||  \\
&\leq  || U_0(t) - e^{-i\Omega_1(T)}|| + || U_i(t) - U_0(t)|| \\
& = O(|A|^2Td^{2+r} ||H_c||^2 \\
& \indent + |A|d^{2+r} ||H_c|| + |A|^2 T^2||H_c||^2 ).
\end{split}
\end{equation*}

Choose $A$ and $T$ such that

\noindent
\begin{equation*}
|A|^2 T^2 ||H_c||^2 >|A|d^{2+r} ||H_c||>|A|^2Td^{2+r}||H_c||^2 
\end{equation*}

\noindent
implying $1 > |A| T ||H_c||$ ensuring the convergence of the Magnus series, $d^{2+r}<|A|T^2||H_c||$  and $T > d^{2+r}$. Such a choice is possible by making $A$ sufficiently small, hence weak driving. Then 

\noindent
\begin{equation*}
||U_i(T)-e^{-i\Omega_1(T)}|| = O(|A|^2 T^2 ||H_c||^2). 
\end{equation*}

Note that $\Omega_1(T)$ is a single qubit Hamiltonian acting on the subspace spanned by $|j\rangle$ and $|m\rangle$. By adjusting $\phi$ one can implement 

\noindent
\begin{equation*}
V_i=e^{-i\frac{A}{2}T | \langle j | H_c | k \rangle| \sigma}
\end{equation*}

\noindent
where $\sigma= \pm \sigma_x, \pm \sigma_y$. Now any $SU(2)$ gate $U_2$ can be decomposed in the form $U_2=e^{-ic_1 \sigma_x}e^{-ic_2 \sigma_y}e^{-ic_3 \sigma_x}$ for some $c_1, c_2, c_3$. This is the Cartan decomposition of $SU(2)$, see for example \cite{KhanejaCartan}. Therefore it takes $O(\frac{3}{|A|T})$ gates to generate any  $U_2$ with error $O(\frac{3}{|A|T} |A|^2 T^2 )=O(3 |A| T )$ since the errors accumulate linearly\cite{NielsenChuang}. An arbitrary unitary $U \in SU(d)$ can be implemented by at most $d(d-1)/2=O(d^2)$ $SU(2)$ rotations \cite{NielsenChuang, KitaevVyalyi}. The total error is then $\epsilon= O(3 |A| T  d^2)$. Therefore it requires a total number of $O(\frac{3 d^2}{|A|T})=O(9d^4 \frac{1}{\epsilon})$ operations to implement any unitary with accuracy $\epsilon$. The gate complexity can be improved to $\text{poly}(d)\text{poly}\log(1/\epsilon)$ by generating SU(2) gates via the Solovay-Kitaev algorithm\cite{Nielsen20063}. The ability to implement any unitary in the interaction picture implies the ability to implement any unitary noting that $U(T')=e^{-i T' H_0}U_i(T')$.

Note that since in our setting we have a drift term 
whose inverse cannot be reached directly we could not invoke the 
discrete Solovay-Kitaev bound \cite{NielsenChuang, Nielsen20063}  
or bounds relating optimal control costs to gate complexity
\cite{Nielsen20061, Nielsen20062}. 

Thus the efficiency of universal quantum computation scales exponentially with the total 
number of spins. Are there architectures enabling control complexity 
to scale polynomially, or sub-exponentially, yet the fraction of controlled 
qubits vanishes with $n$?  One such architecture is to decompose the graph 
into $N$ connected blocks, $b_j$, each containing $L$ qubits with 
boundaries of size $B$ between them where $L>>B$ for large $n$. The 
boundaries are promoted to be controls. And we require that for each block one can apply an additional control term, which together with the Hamiltonian of the block satisfy the assumptions of the complexity result proved above. Thus neighboring blocks are 
separated by a number of controlled qubits. The ability to perform arbitrary 
transformations on control qubits make it possible to completely decouple 
the blocks \cite{Lloyd1999}. Decoupling every block but two, one can 
perform quantum computation on two adjacent blocks with efficiency 
$~\text{s}(2L)$. In an arbitrary network, two blocks are at 
most $N$ blocks apart from each other. To transfer quantum information 
between two arbitrary blocks or equivalently to apply any quantum 
operation to arbitrary two blocks one applies the following procedure. 
Let $b_{p(1)}b_{p(2)}...b_{p(N)}$ be a path of blocks between maximally 
separated $b_{p(1)}$ and $b_{p(N)}$  where $p( \cdot )$ is some permutation 
of N blocks. Quantum information is mediated through the network by first 
decoupling the adjacent blocks $b_{p(1)}$ and $b_{p(2)}$ from the rest 
of network and enacting a quantum transformation on it. Then $b_{p(2)}$ 
and $b_{p(3)}$ are decoupled from the rest and quantum information is 
transferred between these blocks. Continuing this way, quantum information 
can be mediated between any two blocks at most using $O(N)$ pairwise 
decoupling operations. The total gate complexity of applying any quantum 
operation between any two blocks is at most 
$O(N \text{s}(2L))=O(N 2^{2xL} / \epsilon ))$. 
How can the required operations be made to depend sub-exponentially to n? 
Take a family of spin networks, $G(n)$, indexed by the total number of qubits. 
If each network in the family admits a decomposition into $N$ blocks of 
size $L=\text{log}N/2x$ while $n=N\text{log}N/2x$, the complexity can be made 
polynomial as $O(N \text{s}(2L))=O(N^2 / \epsilon ))=O(n^2 / \epsilon)$. 
If also the fraction of controls, $c/n$ can be made to vanish as $n$ grows 
large, where $c$ is the number of controls, one has a scalable quantum computer architecture with a small 
fraction of controls whose gate complexity is sub-exponential. 

Not every family of graphs, $G(n)$, admits a 
decomposition into blocks such that the fraction of controls vanishes 
while the number of elementary operations needed scales polynomially with the 
number of vertices. 
To distinguish between efficiently controllable graphs and
and graphs that are not efficiently controllable, we now present
a formal definition of the efficently controllable family of graphs. 

{\textit{Definition: Efficiently controllable family.}} A family of graphs, 
$G(n)$, indexed by the number of vertices, $n$, is called an efficiently controllable  family if for every $n$ there exists a decomposition 
into connected sub-graphs, blocks, $G(n)=\cup_{K=1}^{N(n)} G_k$ such 
that $\text{lim}_{n\rightarrow \infty} \sum_{1=j<k=N}|G_j \cap G_k| /n \rightarrow 0$ 
where $|G_j \cap G_k|$ is the cardinality of $G_j \cap G_k$, the controls 
between two blocks; in addition, we require that control
complexity 
$D(n) 2^{2xL}/ \epsilon = O(\text{poly}(n),
\text{poly}(1/\epsilon))$ 
where $L(n)$ is the maximum size of blocks 
%$L(n)= \frac{1}{N} \sum_{k=1}^{N(n)} |G_k|$ 
and $D(n)$ is the diameter 
of the graph formed by the blocks. 
%Further we require that 
%$\text{variance}(|G_k|)=O(e^{-n})$ so that each block assumes the same size, 
%the average $L$, rapidly as $n$ grows large. 
Note that the definition can be easily generalized to control of classical networks and other complexity 
measures. 

We give a simple example of a scalable network architecture. Quantum 
information can be transferred from one end to the other of a one dimensional 
chain of $n$ qubits using a fraction of them as controls. This fraction can 
be chosen so that it vanishes as $n$ goes to infinity, and the number of 
elementary operations required scales polynomially with $n$. Assume $N$ 
blocks of qubits of size $L-1$. Between neighboring blocks lies a single 
control qubit. Then the fraction of controls is $c/n=1/L$. We choose 
$L=\text{log}N/2x$ so that $c/n$ vanishes as $n$ goes to infinity. In order 
to enact arbitrary unitary operations between the blocks lying at the right 
and left ends, one first decouples block 1 and 2 from the rest of the 
chain and transfers quantum information coherently from 1 to 2, then 
decouples block 2 and 3, then 3 and 4, etc. Thus it takes $O(N)$ steps 
to mediate quantum information between the blocks that lie at the ends.  
The number of elementary operations required to perform arbitrary 
operations with accuracy $\epsilon$ between adjacent blocks is of the 
order $\text{s}(2L-1)$. Thus the total number of elementary operations needed 
to couple the blocks at the ends
is $O(N\text{s}(2L-1))$. With the choice we made for $L$, the quantum gate 
complexity to enact any desired quantum logic operation between
any two blocks is at most $O(n^2)$.

The previous scheme can be easily generalized to a family of $d$-dimensional 
cubic lattices. We take $N^d$ blocks of size $L^d$ where total number of 
qubits is $n=N^d L^d$. Between two adjacent blocks lies a $d-1$ dimensional 
layer of control qubits. The fraction of controls is again $c/n=1/L$. 
Quantum information can be transferred between blocks lying in the 
opposite diagonal ends by $O(dN)$ pairwise operations on blocks lying 
in the interior. The number of elementary operations required to enact 
quantum logic between adjacent blocks with accuracy $\epsilon$ is given 
by $\text{s}((L-1)^{d-1} (2L-1))$. Choosing 
$L=( \frac{(d-\frac{1}{2})}{2x} \text{log}N )^{\frac{1}{d}}$ 
the total number of elementary quantum operations is given by 
$O(dN\text{s}(2L^d))=O(d n^2 / \epsilon))$ while 
$c/n$ is vanishingly small in the limit of large $n$. 
Now, the generalization to lattices or uniform tilings is evident. 
In the presence of symmetries \cite{symmetry} (existence of a 
subalgebra of the Lie algebra commuting with the drift and all the 
control terms) complete controllability is lost. However one can 
generically break symmetries by perturbations in the coupling Hamiltonian 
or controls. Therefore assuming controllability, efficient controllability 
follows.
%The lattice formed by the unit cells of an arbitrary lattice is cubic. 
%But the number of spins at each unit cell is fixed. Therefore the same 
%arguments show that for any lattice there exists a family of networks 
%labeled by the number of vertices, $G(n)$,  whose members are sub-graphs 
%of the lattice, such that the fraction of controls can be made 
%vanishingly small while universal quantum computation can be done efficiently. 

Note that the construction of efficiently controllable
families given above does not require the dimension $d$ to
be an integer.
%it holds as long as the number of points within graph 
%distance $r$ goes as $r^d$ for any $d\geq 1$.   
Fractals such as the Sierpinski gasket automatically generate
efficiently controllable families.

An efficient way to generate efficiently controllable families is via
site percolation \cite{Coniglio}. Consider an infinite lattice of spins 
where spins sitting in adjacent lattice sites interact with probability 
$p$ and the interaction probabilities for each edge connecting 
lattice sites are independent. When $p$ is just above the percolation 
threshold $p_c$, the graph is connected with unit probability, 
while the structure of the cluster formed is a fractal \cite{Herrmann}. 
At this point, removing a vanishingly small fraction of the spins
at random separates the graph into disconnected pieces: that is, those
removed spins form the interfaces between those pieces of the graph.
The largest size of those disconnected pieces can be estimated as
follows.  Start at $p\approx p_c$ and remove a fraction $\delta$
of the spins.    A group of $N$ previously connected spins will
remain connected if, by a statistical fluctuation, the fraction
of connections within that set remains above $p_c$.  Otherwise,
the group will become disconnected for large $N$.  The average
fluctuation in the number of connections in the group goes
as $\pm \sqrt{p_c(1-p_c) N}$.   The probability that the group
remains connected goes as $e^{-\delta^2 N/p_c(1-p_c)}$.
Accordingly, if one removes a fraction $\delta$ of the spins,
the largest connected group size goes as $O(\delta^{-2})$.
This gives the same scaling for the fraction of control spins
required as that for a two-dimensional lattice, where a group
of size $N$ has a boundary of size $O(\sqrt N)$.  But that
family is efficiently controllable, as shown above.  Consequently
a graph just above the percolation threshold realizes an
efficiently controllable family: universal quantum computation
can be effected by controlling a vanishingly small fraction of
the spins.  The same argument holds for other families of graphs
with percolation thresholds, e.g., Erdos-Renyi graphs \cite{Erdos}.

The site percolation construction above can be applied to scale-free networks 
characterized by the degree distribution $P(k) \sim k^{-\alpha}$ where $P(k)$ 
is the probability for a site to be connected to $k$ other sites. 
For random removal of sites,  the percolation threshold is either $0$ or 
finite \cite{Cohen2000}. However for $\alpha=2$, the removal of high 
degree nodes makes the percolation threshold approach $1$ and removing 
a fraction $ \sim1/N$ of nodes is sufficient to break down the 
network \cite{Cohen2001} into clusters of size 
$\text{log}N/\text{loglog}N$ where $N$ is the total 
number of sites\cite{Aiello2001}. Therefore scale-free networks 
with $\alpha=2$ can be made efficiently controllable. 
Take high degree nodes as controls for decoupling, and take 
one node for each decoupled cluster as the control for enacting quantum gates. 
The total number of controls required to perform quantum computation
efficiently is then a vanishing fraction of total number of sites.

We have exhibited a wide variety of graph families that are
efficiently controllable.    Given a graph family, how hard is
it to determine whether it is efficiently controllable or not?
The problem of finding the minimum set of nodes to control a graph
given a constraint on the maximum block size is a graph partitioning
problem
\cite{Arora, Schaeffer, Fortunato201075}. 
Such problems are generically NP-hard.  Accordingly, we
anticipate that the problem of determining whether a family
of graphs is efficiently controllable is also NP-hard, although we have
no proof.      

The purely graph-theoretic definition of an efficiently controllable
family has applications outside of quantum control theory.
Consider for example the problem of approximating the 
ground state energy of a system, classical or quantum, whose 
interactions correspond to an efficiently controllable graph.
The construction of efficiently controllable graphs shows that
the problem of finding a state whose energy is within a multiplicative
factor $\epsilon$ of the actual ground state energy is polynomial
in the size of the system.   More precisely, consider
a quantum Hamiltonian described by the 
graph $G=(V, E)$, where each vertex corresponds to a variable and each
edge to a pairwise interaction. 
We want to find a state whose energy is within a factor $\epsilon$ of
the actual ground state.
Let $n$ be the number of variables, and
$N$ the number of clusters, each of size $\log N$, so that
$n=N\log N$. 
Disconnect and decouple the clusters of size $\log N$ by removing the 
control qubits, the boundaries between 
the clusters, to get the Hamiltonian $\tilde{H}=\sum_{k}H_{C_k}$ where $H_{C_k}$ is the Hamiltonian 
acting on the cluster $C_k$. The error introduced in calculating the 
ground state energy is at most $\epsilon n
$ where $\epsilon$ is the fraction of controls, i.e. the ground state energy of $H$ is $\epsilon n$ close 
to that of $\tilde{H}$. But the ground state of $\tilde{H}$ is the tensor product of the ground states of $ 
\{ H_{C_k} \}_k$. By standard matrix diagonalization techniques the ground state energy of $H_{C_k}$ 
can be found in $O(\text{poly}~N)=O(\text{poly}~n)$ steps. 
There are $N$ clusters, so it still takes only 
polynomial steps to calculate the ground state energy of 
$\tilde{H}$ and therefore to approximate that 
of $H$ within accuracy $\epsilon N$ which vanishes as $n$ becomes large. Note that our construction is a polynomial time approximation scheme for finding the ground state energy of a 2-local Hamiltonian \cite{Kempe} using clustered product states. Although our construction is in the spirit of product state approximations to ground states \cite{Brandao, Gharibian}, we are approximating with multiplicative error instead of additive error.

This paper investigated the requirements for being able to control
extended systems efficiently.
Quantum systems that can be controlled in
time polynomial in the number of coupled
variables in the system Hamiltonian, by only operating on a vanishingly 
small fraction of those variables, correspond to efficiently controllable
families of interaction graphs.  Such graphs can be divided into
clusters of size $O(\text{poly}(\log n))$ while removing a fraction
$\epsilon$ of the $n$ vertices, with $\epsilon \rightarrow 0$ in
the limit $n\rightarrow \infty$.    
Canonical graph families such as regular lattices are readily shown 
to be also efficiently controllable.  The general criterion for when
families of graphs admit polynomially efficient universal quantum 
computation yet using vanishing fraction of fully controlled qubits 
is an open question. Other open questions include the computational 
complexity of construction of efficiently controllable families and 
whether existing heuristics for graph partitioning problems can 
be exploited to find approximate solutions. 
%Another direction of inquiry 
%is the relation of efficiently controllable families to expansion parameters 
%and therefore to spectral properties of the graph Laplacian \cite{Chung}. 
In this article, several efficiently controllable families were constructed. 
Further open questions include scalable 
architectures in the presence of coupling to environment, 
possible refinements or modifications of the definition 
of efficiently controllable families and their relation to classical 
graph properties.
\bibliographystyle{unsrt}
\bibliography{references}

\begin{thebibliography}{10}
\bibitem{note3}
The new drift Hamiltonian is then $H_0'=H_0+\lambda H_c $. If $H_c$ failed to
  break the degeneracy, the Lie span of $H_0$ and $H_c$ would have failed to
  generate the entire algebra and the system would have not been controllable.
\bibitem{note1}
$P_kH_cP_m\neq 0$ does not need to hold for all $k$ and $m$. For the general
  case, construct the graph $G=(V, E)$ whose vertex $k$ corresponds to $P_k$,
  and there is an edge between vertex $k$ and $m$ only if $P_kH_cP_m\neq 0$.
  Then if $G$ is connected, the system is controllable \cite{Altafini}. The
  gate complexity given below would be increased by a factor of the diameter of
  $G$.
\end{thebibliography}
\end{document}